\documentclass[twocolum, switch]{article} %

\usepackage[utf8]{inputenc}	%
\usepackage[T1]{fontenc}	%

\usepackage{preprint}

\usepackage{xcolor}		%
\usepackage[colorlinks = true,
            linkcolor = black,
            urlcolor  = blue,
            citecolor = cyan,
            anchorcolor = black]{hyperref}

\usepackage[sort&compress,numbers]{natbib}

\usepackage{array}
\usepackage{mathtools}
\usepackage{bm}

\usepackage[nomain,acronym]{glossaries-extra}
\makeglossaries
\setabbreviationstyle[acronym]{long-short}
\newacronym{ai}{AI}{artificial intelligence}
\newacronym{ml}{ML}{machine learning}
\newacronym{llm}{LLM}{large language model}
\newacronym{knn}{KNN}{k-nearest neighbors}
\newacronym{svm}{SVM}{support vector machine}
\newacronym{pfn}{PFN}{prior-data fitted network}
\newacronym{mlp}{MLP}{multi-layer perceptron}
\newacronym{hpo}{HPO}{hyperparameter optimization}
\newacronym{dt}{DT}{decision tree}

\usepackage{siunitx}
\usepackage{booktabs}

\usepackage{tabularx}
\usepackage{makecell}

\usepackage{flushend}

\usepackage{enumitem}

\usepackage{amssymb,amsmath}

\usepackage{soul}

\usepackage{tikz,xcolor,hyperref}

\usepackage{authblk}

\author[1]{Cyril Picard}
\author[1]{Faez Ahmed}

\affil[1]{Department of Mechanical Engineering, Massachusetts Institute of Technology}

\title{Fast and Accurate Zero-Training Classification for Tabular Engineering Data}

\begin{document}

\twocolumn[ %
  \begin{@twocolumnfalse} %

\maketitle

\begin{abstract}
In engineering design, navigating complex decision-making landscapes demands a thorough exploration of the design, performance, and constraint spaces, often impeded by resource-intensive simulations. Data-driven methods can mitigate this challenge by harnessing historical data to delineate feasible domains, accelerate optimization, or evaluate designs. However, the implementation of these methods usually demands machine-learning expertise and multiple trials to choose the right method and hyperparameters. This makes them less accessible for numerous engineering situations. Additionally, there is an inherent trade-off between training speed and accuracy, with faster methods sometimes compromising precision. In our paper, we demonstrate that a recently released general-purpose transformer-based classification model, TabPFN, is both fast and accurate. Notably, it requires no dataset-specific training to assess new tabular data. TabPFN is a Prior-Data Fitted Network, which undergoes a one-time offline training across a broad spectrum of synthetic datasets and performs in-context learning.
We evaluated TabPFN's efficacy across eight engineering design classification problems, contrasting it with seven other algorithms, including a state-of-the-art AutoML method.
For these classification challenges, TabPFN consistently outperforms in speed and accuracy. It is also the most data-efficient and provides the added advantage of being differentiable and giving uncertainty estimates. 
Our findings advocate for the potential of pre-trained models that learn from synthetic data and require no domain-specific tuning to make data-driven engineering design accessible to a broader community and open ways to efficient general-purpose models valid across applications.
Furthermore, we share a benchmark problem set for evaluating new classification algorithms in engineering design and make our code publicly available.

\end{abstract}

\vspace{0.35cm}

  \end{@twocolumnfalse} %
] %

\section{Introduction}

Engineering design is the systematic process of devising solutions to complex problems, often involving the creation of products, systems, or structures through technical specifications and detailed planning. 
Historically, conventional methods and designer intuition led this endeavor. Yet, today's era has brought a new, advanced approach: Data-Driven Engineering Design. By marrying empirical data with advanced analytics, design decisions are optimized based on real-world data and sophisticated computational evaluations. This transition to data-centric methods, though promising, isn't without its hurdles. Evaluations, be they complex experiments or simulations, are sometimes bypassed due to their time-consuming nature, leading to potentially incomplete decisions. Conversely, fast evaluations are often inaccurate. In this landscape, we evaluate if TabPFN, a general-purpose prior-fitted-data classification model developed by Hollmann \textit{et al.}~\cite{hollmann_tabpfn_2023} can be effective in engineering design. Our findings across eight engineering design problems demonstrate that on average, TabPFN is the fastest and most accurate classification model for tabular data-based classification problems. Below we highlight the motivation behind this work.

\paragraph{Importance of Early Evaluation}
The design process often requires synthesizing large amounts of information to make critical decisions. In the early stages, engineers often want to explore the design space to get an understanding of the constraints set by the requirements (\textit{the feasible domain}) and the performance landscape. However, such evaluations can require complex experiments or numerical simulations and may be too time-consuming to be of practical use. As a consequence, engineers too often drop this step and may have to make decisions with partial information.

\paragraph{Potential of Data-Driven Methods}
What if data from previous experiments or past designs were available? An engineer could apply data-driven methods to identify existing patterns in past data and use the resulting model to identify the boundaries of the valid regions of the design space~\cite{malak_using_2010,yoo_bayesian_2021}, speed-up optimization by pre-filtering solutions~\cite{tsai_constraint-handling_2022}, and evaluate design robustness~\cite{massoudi_robust_2022,wiest_robust_2022,caputo_role_2021}. However, current models need to be trained specifically for each application and no clear guidelines exist for selecting the appropriate data-driven method.

In the absence of such guidance, many practitioners find themselves ensnared in a laborious cycle of hit-and-trial or, worse, settling for inadequate model choices that don't optimally leverage available data. 
For each dataset and in adherence to best practices~\cite{sharpe_comparative_2019}, the typical process demands preprocessing the dataset, selecting a data-driven method for the specific application, running a \gls{hpo} to discover the best parameters for the chosen method, training the model with these parameters, and lastly, evaluating its capability in predicting the quality of candidate designs.

\paragraph{Challenges in Current Data-Driven Approaches}
Embarking on the data-driven journey demands both an intricate knowledge of machine learning and a solid grasp of the specific engineering challenge. This dual expertise is crucial for effective model selection and the configuration of the \gls{hpo}. 
It can be very time-consuming and represent a challenge, especially if new data points are added iteratively, e.g., in an active learning scheme to identify the boundaries of the feasible space~\cite{chen_active_2018}. Further, without a large and diverse dataset, training a classical ML model from scratch can lead to overfitting~\cite{li_assembly_2022}, where the model learns to perform well on the training data but poorly on new, unseen data. In engineering design, small datasets are unfortunately a common situation~\cite{regenwetter_deep_2022}.

\paragraph{Previous Solutions and Their Shortcomings}
Different attempts have been made in past years to tackle part of these issues. Ensembling approaches, which combine several data-driven methods or hyperparameter sets, reduce the dependence on choosing a single set of methods and hyperparameters valid across the whole domain (e.g., XGBoost~\cite{chen_xgboost_2016}). 

Another popular solution is automated machine learning (AutoML). Tools like AutoGluon~\cite{erickson_autogluon-tabular_2020} are examples of this, turning the whole process into one simple algorithm. They handle everything from cleaning up data to picking the right methods and then combining the results. AutoML tools have shown great results in regression and classification tasks~\cite{erickson_autogluon-tabular_2020}. In engineering design, they also did well in tasks like designing bicycle frames~\cite{regenwetter_framed_2023}. While they often give accurate results and make it easier for more people to use advanced methods, they can be very slow to train and need to be trained again for each new dataset. Others have investigated the use of data-augmentation techniques to alleviate the challenges of small datasets and have shown that their models trained on ``pseudo'' or ``fake'' data could have increased performance~\cite{du_generating_2021,li_assembly_2022}.

\paragraph{Potential of Pre-trained Transformers}
Imagine if we didn't have to train a model for every specific dataset but instead had general-purpose model that knows a lot and can quickly adapt to new tasks without any new training. This isn't a far-fetched idea. In the world of natural language processing, pre-trained models, in the form of large language models (LLMs), have already shown they can do this. For instance, the Text-to-Text Transfer Transformer (T5) learns various text tasks, like translating languages or summarizing articles, and then uses this knowledge on new text~\cite{raffel_exploring_2020}.

Most approaches use the transformer architecture~\cite{vaswani_attention_2017}, a large neural network with an attention mechanism that changes the importance of the various parts of the input based on the input (called context) itself. Transformers are hard to train due to their size (millions of parameters) and data requirements (millions--trillions of data points). Once trained, however, they can be fine-tuned with much less effort (e.g., LoRA~\cite{hu_lora_2022}) and some are even ready for use out-of-the-box and can be easily integrated into applications for quick deployment. Indeed, their architecture enables \textit{in-context learning}: Show them a few examples of a new task at inference time---a sequence of input, expected output---and let them subsequently apply it~\cite{li_transformers_2023}. Many readers may have used pre-trained general-purpose models and observed in-context learning in action when using tools like ChatGPT, Bard, or Claude. These models and tools have shown immense promise in language and vision tasks.

\paragraph{Potential of PFNs for Engineering Design}
In contrast, in engineering design, we often deal with data arranged in tables or matrices, where each row could represent a unique design and columns might indicate different parameters or performance metrics. Historically, deep learning hasn't been the best fit for such tabular data, especially when compared to other machine learning techniques~\cite{shwartz-ziv_tabular_2022}. 
However, the trend might be changing. Recent studies have been exploring how transformer models can be effectively applied to tabular data~\cite{zhu_xtab_2023,muller_transformers_2022,hollmann_tabpfn_2023}.

In particular, \glspl{pfn} learn to approximate a large set of posteriors by being trained on millions of dynamically generated synthetic datasets sampled from a prior inspired by structural causal models. They can then make well-calibrated probabilistic predictions for new datasets at inference time in a single forward pass. They are fast and accurate at approximating Bayesian inference~\cite{muller_transformers_2022} and at classification on various \gls{ml} benchmark~\cite{hollmann_tabpfn_2023}.
Considering they are specifically geared towards smaller datasets, \glspl{pfn} have the potential to revolutionize data-driven tasks in engineering design, especially if they prove their efficacy on datasets that are specific to the field.  In this paper, we show that on average, they are not only the fastest and most accurate method across eight different engineering problems, but also provide two desirable and valuable properties: differentiable and well-calibrated uncertainty estimations.

\paragraph{Goal and contributions}
The core of our study centers on evaluating if the recently publish PFN for classification, named TabPFN~\cite{hollmann_tabpfn_2023}, which has been trained only on synthetic data and have never seen \textit{airfoils}, \textit{beams}, or \textit{bicycle frames}, can still be effective on such engineering design problems. Our objective is to illustrate that general-purpose models that negate the need for training or fine-tuning on new datasets, can be effective and even superior to leading-edge AutoML and traditional algorithms in terms of speed, accuracy and data efficiency, especially in the domain of engineering design.  
This would mark a significant departure from conventional data-driven approaches, since such models can be readily used out-of-the-box, and do not require specialized knowledge or large datasets for training.
Specifically, the contributions are:
\begin{enumerate}
    \item Enhanced Benchmarking Tools: We introduce a collection of eight engineering design datasets on airfoils, bicycles, solar heat exchangers, trusses, and welded beams, tailored for benchmarking classification algorithms.
    \item Accuracy and Speed Comparisons: By performing a thorough comparative study of TabPFN's efficiency and speed against common classification algorithms and state-of-the-art AutoML methods using these datasets, we show that TabPFN is the highest ranked method.
    \item Efficiency of Data Utilization: We provide a metric to compare classification methods on how efficiently they use available data and show that TabPFN is most efficient in using data.
    \item Insight into PFNs' Promise: Beyond just quantitative results, we engage in a reflective discourse on the latent potential of \glspl{pfn}, training on synthetic data and in-context learning within the engineering design domain. 
    We discuss important factors beyond speed and accuracy for selecting classification models, such as interpretability, updatability, and uncertainty quantification.
    \item Open-Source Code and Dataset: In the spirit of fostering collaborative advancement, we are also releasing our eight datasets and the associated codebase at \url{https://decode.mit.edu/projects/pfns4ed/}\footnote{Data and code will be released upon paper acceptance.}. This will act as a springboard for other researchers to delve deeper, utilize, and further the research in engineering classification.

\end{enumerate}

\paragraph{Outline}
A general background on the use of classification in engineering design and classification algorithms is provided in Section 2. Section 3 details the considered algorithms and the methodology applied to compare the algorithms. The results of the performance-speed comparison and the data-efficiency analysis are presented in Section 4. Finally, Section 5, discusses the impact of the strong results of TabPFN on design, the limitations, and the outlook for \glspl{pfn} at large.

\begin{figure*}[ht]
    \centering
    \includegraphics{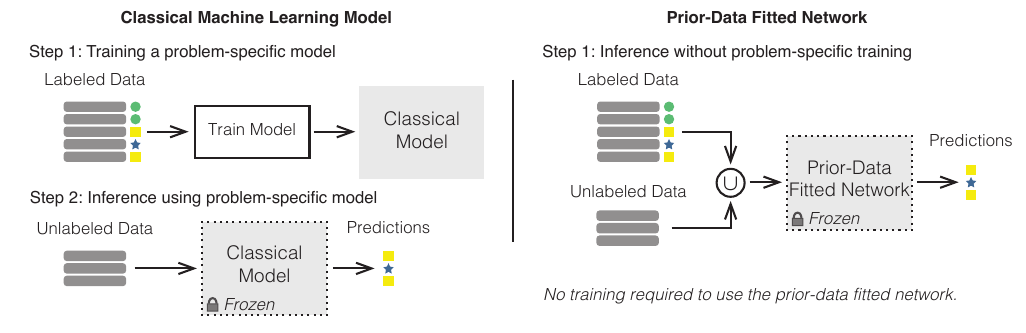}
    \caption{The conceptual difference between classical classification models and prior-data fitted networks, showing how PFNs do not require a training step, leading to fast predictions.}
    \label{fig:classification-overview}
\end{figure*}

\section{Background}
In this section, we formally introduce the classification task, and provide the background into classical machine-learning classification methods. We further highlight the differences with \glspl{pfn}, which are summarized in Fig.~\ref{fig:classification-overview}.

\subsection{Classification and Its Use in Engineering Design}

In general terms, classification is the task of identifying which class \(c\) from a predefined set of classes \(\mathcal{C}\) a candidate design \(\mathrm{x}\) belongs to.  The task is called binary classification when there are two classes, often denoted as 0 (negative) and 1 (positive). With three or more classes, the problem is referred to as multi-class classification, where classes may or may not have an ordinal relation. In engineering design, the classes can represent a wide array of concepts from bike categories (road, MTB, gravel,...)~\cite{regenwetter_biked_2021}, to product ratings~\cite{singh_machine_2017}, to performance and constraint-satisfaction indicators (valid or invalid design)~\cite{yoo_bayesian_2021,wiest_robust_2022,massoudi_robust_2022}. The latter, in particular, is common in engineering as it allows to easily map regions of interest for design. It can be mathematically defined for the binary case by:
\begin{equation}
    c_x = \begin{cases}
    1,& \text{if } f(x)\leq f_{thresh} \text{ and } g(x) \leq 0 \\
    0,              & \text{otherwise}
\end{cases}
\end{equation}
Where $f(x)$ represents a performance metric that should be lower than a threshold $f_{thresh}$ (e.g., total weight $\leq 5$ kg for a bike frame) and $g(x)$ is a set of inequality constraints to be satisfied.
Noteworthy, multi-class classification problems are commonly more difficult due to the increase in prediction dimensionality.

When obtaining the true class is impractical, such as when it is derived from expensive numerical simulations or it requires the collection of human feedback, data-driven methods can be used as faster surrogates. Following the above definition, a fast classifier is convenient to perform design exploration to identify regions of promising designs, speed-up optimizations, and evaluate the robustness of designs. In some cases, the identified designs are then evaluated with the expensive functions and they become new data points that can be added to the training dataset, and the classification model needs to be updated iteratively. 

\subsection{Classical Machine-Learning Classification Methods}\label{sec:existing_methods}
The use of typical machine-learning classification models generally involves a two-step process: training followed by inference. In the training step, a labeled dataset is provided. Then a model and training approach are selected along with their parameters, called hyperparameters. Hyperparameters are parameters that are not learned from the data but are set by the user before training the model. The training step results in a learned internal problem representation, known as ``the model'', which depends on the selected method. Examples include weights and coefficients of basis functions (support vector machine), branching rules (decision trees), or proximity trees (K-D trees for nearest-neighbor searches). In the inference step, the representation is frozen and used to predict the labels of unlabeled data points.

The key challenge with this approach lies in the need to choose the appropriate method and its hyperparameters~\cite{sharpe_comparative_2019}. Indeed, each modeling method has limitations and requirements that stem from the underlying assumptions. As a consequence, the training step may need to be repeated multiple times to find the best combination of model and hyperparameters for a given task.

Without careful consideration, applying this optimization step can rapidly lead to overfitting: A Model becomes too closely tailored to the training data and loses its ability to make accurate predictions on unseen data. To mitigate this risk, cross-validation is used to assess the performance of ML models. The training dataset is divided into multiple subsets, or folds, where each fold is used as both a training set and a validation set. The model is trained on the training set and evaluated on the validation set. This process is repeated multiple times, with different folds used as the validation set each time. The average model performance across folds is then used to guide the hyperparameter optimization.

To go beyond the modeling limitations of a single model type, ensembles can be built. Ensembles combine multiple models, often of different types or trained on different subsets of the data, to make predictions. By aggregating the predictions of individual models, ensembles can often achieve better performance than any single model~\cite{rokach_ensemble-based_2010}. While effective, all these steps come at a large computational expense, especially during the training step. As such, the process needs to be repeated for each new dataset.

\subsection{Prior-data Fitted Networks}\label{sec:pfn}
\begin{figure*}
    \centering
    \includegraphics[width=\linewidth]{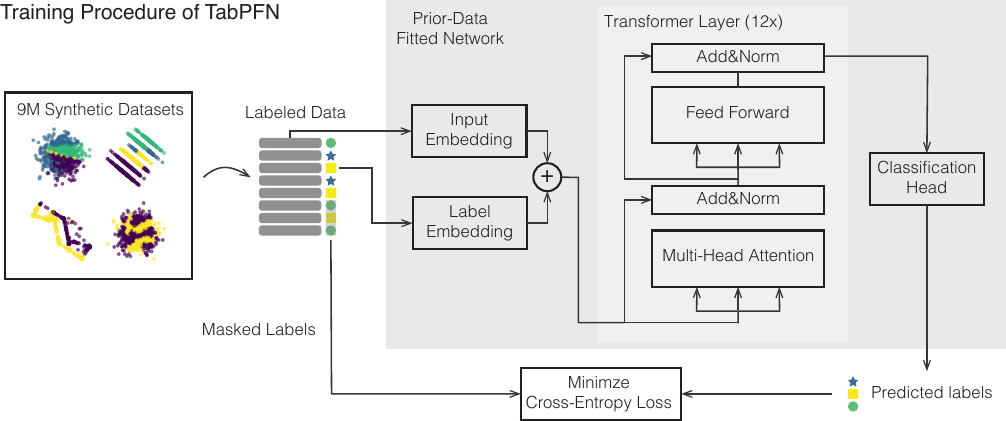}
    \caption{Overview of the training procedure of TabPFN enabling it to learn the classification algorithm in general. A each training step, a dataset is sampled from a pool of synthetic datasets and arbitrarily cut into a train and test. The labels from the test set are removed and are used to calculate the cross-entropy loss used as training metric.}
    \label{fig:tabpfn}
\end{figure*}

Contrary to classical methods, \glspl{pfn} leverages ideas from causality and Bayesian learning to meta-learn a powerful inductive bias for fast predictions on new tabular data. They are trained to learn an ``algorithm'' or general task and are thus not restricted to a given problem and dataset. Considering for example TabPFN~\cite{hollmann_tabpfn_2023}, it has been trained to perform \textit{data-driven classification}. TabPFN is thus a general-purpose classification algorithm. To apply it to a new classification problem, both training samples (pairs of feature vectors and labels) and unlabeled samples to be evaluated are passed to the model simultaneously. While the points are called ``training'' samples, no training or fine-tuning is performed: The samples undergo a single forward pass through a frozen model.\footnote{A frozen model is a model whose parameters, weights and biases are not changed.} Instead, the ``training'' samples condition the model through its attention mechanism to use the prior that matches the new problem most to predict the labels of the unlabeled samples, very much like the text prompts to LLMs conditions their responses.

In terms of architecture, TabPFN~\cite{hollmann_tabpfn_2023} uses an encoder-only transformer architecture~\cite{vaswani_attention_2017} without positional encoding, see Fig.~\ref{fig:tabpfn}. Each feature vector and label is encoded as a token, summed together and passed to the sequence of transformer layers. In the attention head, training samples can attend to each other, while unlabeled samples can only attend to training samples. This attention mechanism enables conditioning the model to new datasets at inference time. Finally, the output of the last transformer layer is passed through a simple neural network to output the classification probabilities.

Like LLMs that accept prompts with different numbers of words, TabPFN does not expect the train and test sets to have a fixed number of rows. For example, it is possible to feed it with \num{100} training points and ask predictions for thousands of test points, or {\num{2000}} training points and only one test point. However, unlike LLMs where the order of the words is important, the lack of positional encoding in TabPFN makes it invariant to permutations of the input.

In terms of training, TabPFN needs to be trained only once and the procedure is summarized in Fig.~\ref{fig:tabpfn}. Specifically, the available model has been trained on more than nine million diverse datasets. At each training step, a new dataset is sampled and arbitrarily cut into a train and test set of various length. The model is then asked to predict the masked labels from the test set. Using the real and predicted labels, a cross-entropy loss is calculated and used to update the model to minimize this loss.

Learning a general algorithm like this works under the hypothesis that the model has been trained on a diverse set of tasks and that the application datasets resemble the prior used for training. Indeed, TabPFN has been trained on more than nine million diverse datasets, and can thus use its large ``past experience'' to make better predictions, especially for small or biased datasets. There are however limitations. The synthetic datasets only include problems with one to 100 features and two to ten classes. Feature subsampling and building ensembles can provide a way around those limitations.  Also, while its architecture lends itself to the parallel execution capabilities of GPUs, transformers are memory-intensive with a quadratic dependency on input length. Consequently, the number of points that can be processed by a \gls{pfn} is memory-bounded. On current consumer GPUs, TabPFN can handle up to about 5000 training points.

\section{Methods}

This section details the selected datasets and the conditions under which the different classification methods have been compared, their parameters, and the analysis methods.

\subsection{Engineering Datasets}

To start, the selected problems from engineering design and related datasets are discussed. We gathered a set of eight problems from the literature~{\cite{sharpe_comparative_2019,heyrani_nobari_pcdgan_2021,regenwetter_biked_2021}}. The goal was to have a diverse set of problems with different number of features (2--100), binary and multi-class, and with and without discrete variables. The eight problems are summarized in Table~\ref{tab:datasets}.

\begin{table*}[t]
    \centering
    \caption{List of selected datasets along with key metrics. Problems that have a separate test set are indicated with an $^\ast$.}
    \label{tab:datasets}
    \begin{tabular}{@{}llccclrr@{}l@{}}
      Dataset                  & Modality    & Features & Discrete Var & \multicolumn{2}{l}{Classes and their proportion} & Train size & Test size       \\ \midrule
      Airfoil  binary    & Coordinates & 100      & \texttimes   & 2 & 65\% / 35\% & 886 &  222 \\
      Airfoil multi-class & Coordinates & 100      & \texttimes   & 4 &  47\% / 25\%  / 19\% / 9\%    & 886         & 222       \\
      FRAMED validity           & Parameters  & 39       & \checkmark   & 2 & 91\% / 9\% & 3609 & 903 \\
      FRAMED safety             & Parameters  & 39       & \checkmark   & 4 &  49\% / 31\% / 15\% \ 5\%    & 3236  & 810  \\
      Solar HEX      & Parameters  & 2        & \checkmark   & 2 & 52\% / 48\% & 500  & 2500 &$^\ast$  \\
      Three-bar truss               & Parameters  & 6        & \checkmark   & 2 & 66\% / 34\% & 1000  & 5000&$^\ast$ \\
      Welded beam               & Parameters  & 4        & \texttimes   & 2 & 98\% / 2\% & 2000 & 2500&$^\ast$ \\
      Welded beam (+)    & Parameters  & 4        & \texttimes   & 2 & 84\% / 16\% & 2500 & 2500&$^\ast$ \\
    \end{tabular}
    
\end{table*}

The study incorporates diverse datasets, representative of various facets of engineering design problems. A portion of these datasets has been chosen to facilitate a comparison with earlier works, particularly based on the review by Sharpe et al.~\cite{sharpe_comparative_2019}. Specifically, three realistic design problems from the said review were chosen---the three-bar truss, the solar heat exchanger (solar HEX), and the welded beam design problems---owing to the availability of data or the possibility of its evaluation. In addition, the FRAMED dataset~\cite{regenwetter_framed_2023} and an airfoil dataset inspired by~\cite{heyrani_nobari_pcdgan_2021} are considered to evaluate the classification methods on problems with higher dimensionality.

These design problems all revolve around identifying designs that not only satisfy certain constraints but also achieve a set minimum performance threshold, which makes them suitable for classification tasks. Some contain discrete variables, meaning only specific integer values are permissible for those features.

More specifically:

\paragraph{Airfoil Binary and Multi-class Datasets} Originating from the UIUC Airfoil Coordinates database\footnote{\url{https://m-selig.ae.illinois.edu/ads/coord_database.html}}, they consist of the (x,y)-coordinates of 50 points interpolated along the airfoil profiles following~\cite{heyrani_nobari_pcdgan_2021}. Their maximum lift coefficient is calculated using Xfoil~\cite{drela_xfoil_1989} by sweeping the angle of attack between $\SI{-5}{\degree}$ and \SI{30}{\degree}. The binary dataset separates airfoils with a lift coefficient greater than 1.7 and has a class distribution of 65\% to 35\%. In contrast, the multi-class version involves four distinct classes distributed as 47\%, 25\%, 19\%, and 9\% associated with maximum lift coefficients greater than 1.0, 1.5, and 1.8, respectively. There is no predefined test set.

\paragraph{FRAMED Validity and Safety Datasets} These are characterized by 39 parameters, encompassing discrete variables, that characterize bicycle frames. The validity dataset distinguishes between geometrically valid and invalid frames (91\% to 9\% split). The safety dataset presents a multi-class challenge as it separates between frames that fail on both loading tests (49\%), pass only the loading test 1 (31\%), pass only the loading test 2 (15\%), and pass both loading tests (5\%). The geometric constraints and loading cases are further detailed in~\cite{regenwetter_framed_2023}. There is no predefined test set.

\paragraph{Solar HEX Dataset} This dataset contains solar heat exchange designs characterized by two parameters. The label indicates if the exchanger is high-performing and feasible. This is a binary classification problem with a balanced class distribution. Train and test datasets were methodically generated using Sobol and Halton sequences respectively, and subsequently labeled using the codes provided by the original authors~\cite{sharpe_comparative_2019}.

\paragraph{Three-bar Truss Dataset} It consists of six parameters defining the cross-section and material for each bar of a three-bar truss problem. It aims at binary classification, with moderately imbalanced class distribution. Separate train and test sets were provided directly by the original authors. Details about the truss design are available in the supplementary material of the original publication~\cite{sharpe_comparative_2019}.

\paragraph{Welded Beam Datasets} They are two variations that differ only by the method used to sample points. The underlying problem is represented by four parameters and results in a binary classification problem. The Welded Beam train and test datasets were methodically generated using Sobol and Halton sequences respectively, and subsequently labeled using the codes provided by the original authors~\cite{sharpe_comparative_2019}. This sampling results in a highly imbalanced dataset for the welded beam problem. With only 2\% feasible designs, the classification task is very challenging. Consequently, a second variant, named Welded Beam (+) is created by sampling, for the train set, an additional 500 points using Bayesian optimization and the straddle acquisition function~\cite{bryan_active_2005}. This results in an increase to 16\% feasible data points. The test set for this variant is obtained by sampling \num{20000} points from the Halton sequence and then randomly subsampling it to \num{2500} by giving more weight to feasible points.

\subsubsection*{Notes On Preprocessing}

Datasets are considered to be numerical only and no special treatment of discrete variables is done. For other datasets, one-hot encoding or ordinal mapping may be needed. TabPFN and AutoGluon apply their own preprocessing, as such the datasets are provided as is. For all other methods, a standard scaler ($z=(x - \mu) / \sigma$) is fitted and applied to the training data and applied to test data without refitting. Likewise, classes in the training data are mapped to a contiguous list from 0 to the maximum number of classes, and the labels predicted by the models are mapped back to the original space. As these processes are part of TabPFN and AutoGluon, the time required for the preprocessing is also accounted for in the training time for other methods.

\subsection{Classification Methods}
This section covers the classification methods considered in this work. Table~\ref{tab:classification_methods} provides an overview of the selected methods.  We picked three baseline models: k-nearest neighbors, support vector machines, and decision trees. Each represents a different approach to classification and they are known for their simplicity and widespread application in various domains, including engineering design. On top of that, we include methods from two state-of-the-art approaches: gradient-boosted decision trees, and AutoML~\cite{shwartz-ziv_tabular_2022,erickson_autogluon-tabular_2020}. It is worth noting that multi-layer perceptions (artificial neural networks) are not considered, as they are expensive to train and are underperforming on tabular data classification, especially for smaller datasets~\cite{shwartz-ziv_tabular_2022}. In the following, we cover the key components of each method.

\begin{table}[t]
    \centering
    \caption{Overview of the classification methods used and the chosen settings.}
    \label{tab:classification_methods}
    \begin{tabular}{@{}lcl@{}}
                &  GPU  & Settings\\ \midrule
       TabPFN    & \checkmark & Default \\
       XGBoost & \texttimes & Default \\
       AutoGluon$^+$ & \checkmark & \textit{Best quality} preset \\
       AutoGluon$^\sim$ & \checkmark & \textit{Medium quality} preset \\
       XGBoost$^\ast$ & \texttimes & Parameters tuned with SMAC \\
       SVM$^\ast$     & \texttimes & Parameters tuned with SMAC \\
       Decision Trees$^\ast$ & \texttimes & Parameters tuned with SMAC \\
       KNN$^\ast$ & \texttimes & Parameters tuned with SMAC \\
    \end{tabular}
    
\end{table}

\subsubsection{Baseline Models}
For our research, these models were implemented using the scikit-learn library~\cite{pedregosa_scikit-learn_2011}.

\paragraph{k-Nearest Neighbors (KNN)} A non-parametric method, KNN operates on the principle of feature similarity. By considering the $k$ most similar instances from the training dataset, it predicts the classification of new data points. Its simplicity lies in its ability to make decisions based on the majority class of its neighbors, making it intuitive and easy to implement.

\paragraph{Support Vector Machine (SVM)} SVM is a supervised machine learning model that seeks to find the optimal hyperplane that best separates the data into classes. It's especially effective in high-dimensional spaces and is known for its kernel trick, enabling it to create non-linear decision boundaries.

\paragraph{Decision Trees (DT)} As the name suggests, Decision Trees divide the dataset into subsets using a tree-like model of decisions. They recursively split the data based on feature values, making them particularly interpretable, as the decisions can be visualized and understood easily.

\subsubsection{Gradient-Boosted Decision Trees}
Gradient-Boosted Decision Trees (GBDT) is an ensemble technique that builds upon the principle of boosting. It combines the output of multiple shallow decision trees, added iteratively, to improve predictive accuracy and reduce overfitting. At each stage, a new tree is fit to the negative gradient (hence, ``gradient-boosted'') of the loss function, which corrects the mistakes of the previous trees. Over several iterations, this method effectively tunes the model to the intricacies of the dataset, enabling it to handle complex non-linear patterns. One of the most popular implementations of GBDT is XGBoost~\cite{chen_xgboost_2016}, which is renowned for its computational efficiency and capability to handle large datasets.

\subsubsection{AutoML (AutoGluon)}
Automated Machine Learning (AutoML) refers to the process of automating the end-to-end process of machine learning. From data preprocessing, feature selection, and model selection, to hyperparameter tuning, AutoML seeks to produce high-quality machine-learning models with minimal manual intervention. AutoGluon, in particular, is an open-source AutoML library that offers automated model selection and hyperparameter optimization for a variety of machine-learning tasks~\cite{erickson_autogluon-tabular_2020}. It has been shown to work well on the FRAMED dataset~\cite{regenwetter_framed_2023}. AutoGluon is, however, very time-consuming to run. Therefore, we consider two presents: \textit{medium} and \textit{best}. The presets change the number of configurations tested in the \gls{hpo} search and the number of models in the weighted ensemble, influencing both the training and the inference time. 

\subsubsection{Prior-data Fitted Network (TabPFN)}
As described in  Section~{\ref{sec:pfn}}, TabPFN is a novel model tailored for tabular data. Unlike traditional deep learning models that require application-specific models, extensive data and training time, TabPFN capitalizes on the information from previously trained datasets. By leveraging prior knowledge and transferring it to relevant tasks, TabPFN is a general-purpose classification algorithm that offers an efficient alternative for situations where data might be limited or where computational resources are constrained. We use the exact model published in~\mbox{\cite{hollmann_tabpfn_2023}} across all the evaluated problems without any fine-tuning.

\subsubsection{Hyperparameter Tuning}
Following recommended best practices, the hyperparameters of KNN, SVM, DT, and XGBoost are tuned~\cite{sharpe_comparative_2019}. We employed the SMAC3 framework~\cite{lindauer_smac3_2022} (version 2.0.0) for hyperparameter tuning. SMAC3 is a state-of-the-art tool that combines Bayesian optimization and random forest regression to search the hyperparameter space efficiently. To utilize SMAC3, we first defined the hyperparameters and their respective ranges for each method. We employed a 5-fold cross-validation setup, where the train set was split into five subsets, and the models were trained and evaluated on different combinations of these subsets. The average model accuracy across the five folds was the metric to optimize. By iteratively running a set of configurations and evaluating their performance using the accuracy metric, SMAC3 adaptively selected promising hyperparameter settings to explore. This process continued until convergence or a predefined budget was reached (a maximum of 100 configurations). Through SMAC3, we aimed to identify the optimal hyperparameter configuration that maximized the accuracy of our machine learning models, enhancing their predictive accuracy and generalization capabilities. The hyperparameters for each method can be found in the code associated with this work.\footnote{\url{https://decode.mit.edu/projects/pfns4ed/}}

Hyperparameter tuning (including AutoGluon) was run from scratch for each dataset size and random split, as we wanted to mimic a real-world scenario where train data is the only thing that is known and test data are actual points of interest to be evaluated.

\subsubsection{Computational Environment}
The study was conducted on a single computer with a 12th Gen Intel\textregistered~Core\texttrademark~i9-12900K CPU and an Nvidia 3090Ti GPU. We used Python (version 3.10.9) along with scikit-learn (version 1.2.2), XGBoost (version 1.7.4), AutoGluon (version 0.7.0), and TabPFN (version 0.1.8). More details about the Python environment are provided with the code. The use of GPU or CPU was decided based on the methods' compatibility with GPU and trials to identify which hardware was the fastest.
Overall, Table~\ref{tab:classification_methods} provides an overview of the considered methods for classification.

\subsection{Evaluation Protocol}
\subsubsection{Data Handling} 

To control for the stochastic components of each classification method and to rule out lucky data splits, we repeat each training and evaluation step 20 times for each dataset and each method. We start by generating 20 different random permutations of each dataset. For the welded beam dataset, we ensure that at least two feasible samples are present within the first 100 points given its high imbalance. For the Airfoil and FRAMED datasets, the data is further split with the first 20\% becoming the test set, and the last 80\% the train set.

Within each data split, each method is evaluated considering only the first 5\%, 10\%, 20\%, 30\%, 40\%, 50\%, 60\%, 70\%, 80\%, 90\% and 100\% of the resulting training set. This yields a total of 220 train/test executions per dataset per method. With this approach, the change in performance of each method as a function of data quantity can be assessed. In particular, it offers insights into the data efficiency of each method.

\subsubsection{Evaluation Metrics}
In this section, we discuss our choice of evaluation metrics and their respective limitations.

\paragraph{Accuracy} One of the most intuitive metrics in classification problems is accuracy. It measures the ratio of correctly predicted instances to the total number of instances. While accuracy might seem like an obvious choice due to its simplicity, it is not without its flaws, especially when dealing with imbalanced datasets. In such cases, even a naive model that predicts the majority class for all inputs can achieve a high accuracy, rendering the metric misleading. As most of the datasets we study are imbalanced (see Table~\ref{tab:datasets}), we use the F1-score metric for our analysis.

\paragraph{F1-score}
To combat the limitations of accuracy, especially in scenarios with imbalanced classes, the F1-score is often employed. Defined mathematically as:

\begin{equation}\label{eq:f1}
F_1 = \frac{2 \times \text{ true positives}}{2 \times \text{ true positives}+\text{false positives} + \text{false negatives}}
\end{equation}

The $F_1$-score provides a balance between precision (the ratio of correctly predicted positive observations to the total predicted positives) and recall (the ratio of correctly predicted positive observations to all observations in actual class). It is particularly useful when false positives and false negatives have differing costs.

\paragraph{Multi-class $F_1$-Score (Macro Averaging)} The traditional $F_1$-score is intrinsically binary. However, for problems involving multiple classes (such as Airfoil multi-class and FRAMED safety datasets), an adaptation is required. One common strategy is macro averaging. Here, the $F_1$-score is calculated independently for each class and then the average is taken. 

\begin{figure}
    \centering
    \includegraphics[width=1\linewidth]{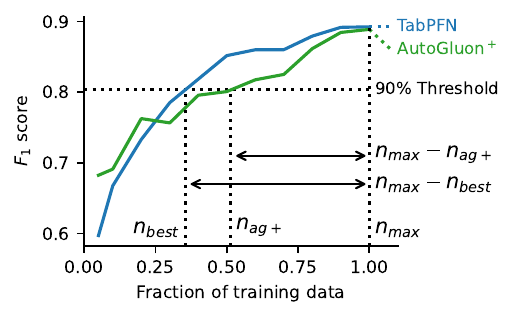}
    \caption{Graphical representation of the suggested data efficiency showcased using two methods for simplicity. Data efficiency is defined as the ratio of how much data is left once a method crosses a performance threshold over how much data is left for the first method to cross the same threshold. In this example, TabPFN and AutoGluon$^+$ have a data efficiency of  100\% and 75\%, respectively.}
    \label{fig:data_eff_def}
\end{figure}

\paragraph{Relative data efficiency}
We evaluate the data efficiency of each method by considering how many more data points are needed to reach some performance threshold  compared to the 
first method reaching the same threshold within a data split. In this work, the threshold is set to 90\% of the best performance achieved across all methods for the given data split. This level is chosen to represent a reasonable classifier. The method is illustrated in Fig.~\ref{fig:data_eff_def}. Mathematically, the efficiency of method $m$ for a given data split is given by:

\begin{equation}\label{eq:data-efficiency}
    \eta_{D,m} = \frac{n_{max} - n_m}{n_{max} - n_{best}}
\end{equation}
Where $n_{max}$ is the total number of points in the dataset, $n_{best}$ is the number of points needed for the best method to reach the threshold and $n_m$ is the number of points needed for the method $m$ to reach that threshold (or $n_{max}$ if it does not achieve the threshold). Using this definition, the first method to reach the threshold is awarded a 100\% efficiency while a method using all data points gets 0\%.

\subsubsection{Statistical Testing Protocol}

Ensuring the validity and statistical significance of our results is pivotal for drawing reliable conclusions. Given the multiple machine learning methods and datasets involved in our analysis, we employed a robust statistical testing protocol to compare the performances of these methods.

\paragraph{Friedman Ranking Test}
This non-parametric test is chosen to assess if there are differences in ranks among multiple paired samples. It is used to identify if there are any differences that would warrant further analysis.

\paragraph{Wilcoxon Significance Analysis}
This non-parametric test, widely recognized for comparing two paired samples, is chosen to assess the difference in ranks of various machine learning methods across our datasets. Unlike the t-test, the Wilcoxon test does not assume a normal distribution and is therefore more versatile, especially when the data distribution is unknown or not normally distributed.

\paragraph{Holm’s Adjustment}
Given the multiple comparisons being made (due to the variety of machine learning methods), there arises the risk of observing a statistically significant difference purely by chance. The issue of multiple comparisons could inflate our Type I error. To control the family-wise error rate and counteract this problem, we employ Holm’s adjustment.

In our study, the protocol involves the following steps:

\begin{enumerate}
    \item Ranking of Methods: For each dataset and data split, the machine learning methods are ranked based on their performance, with the best-performing method receiving the lowest rank.
    \item The Friedman ranking test is applied on the ranks to identify if there are differences between any methods.
    \item If the calculated p-value from the Friedman ranking test is below the significance threshold set at 0.05. All methods are compared in a pairwise fashion using the Wilcoxon significance analysis and Holm's adjustment is used to account for the multiple comparisons.
    \item The final step of our statistical analysis involves interpreting the corrected pairwise p-values to find the classification methods that are significantly different from the others, using the same 0.05 threshold.
\end{enumerate}

Through this rigorous statistical testing protocol, we can confidently infer differences in the performance of the machine learning methods across the datasets, ensuring that our conclusions are not the mere result of chance or the byproduct of multiple comparisons.

\section{Results: A Comparative Study of Classification Methods}
The results of our study are a set of 1760 different train sets evaluated across eight methods. We decompose these results by considering first the performance of all methods of the full datasets. Second, we investigate the effect of dataset size on performance and use the performance change with dataset size to quantify the data efficiency of each method. Third, we assess the overall performance.  Finally, we put these results into perspective considering both performance and speed to nail the advantages of TabPFN.

\subsection{Performance Comparison on Full Datasets}
\begin{figure*}
    \centering
    \includegraphics[width=\linewidth]{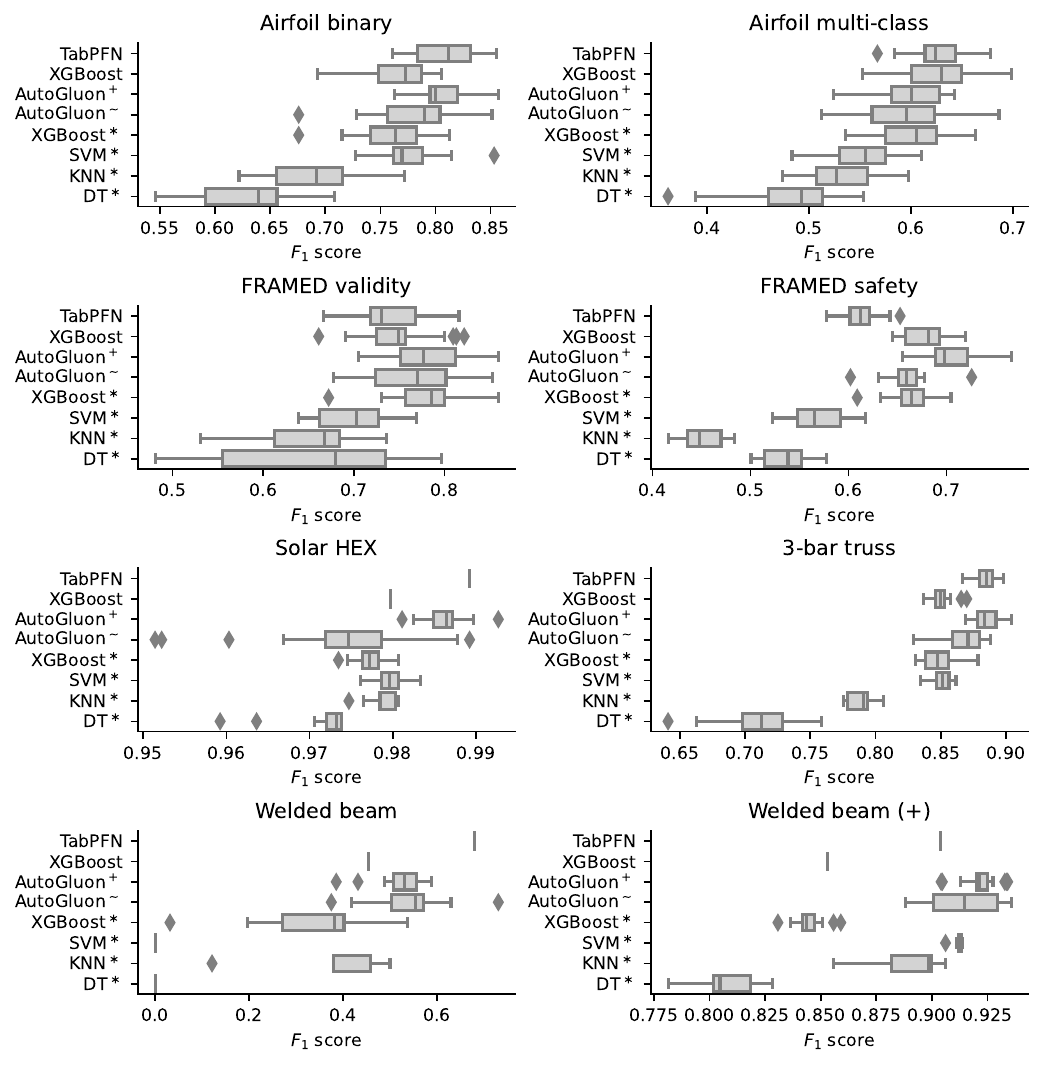}
    \caption{Boxplots comparing the $F_1$ scores of each method for each full dataset. Notice that the $F_1$ score scale is different for each dataset. We observe that TabPFN has the highest median performance in 4 out of 8 problems.}
    \label{fig:f1_boxplot_full}
\end{figure*}

As a starter, we compare the classification performance through the $F_1$ score of all eight methods when considering the full datasets. Figure~\ref{fig:f1_boxplot_full} displays the distribution of performance across the 20 different splits for all methods and datasets. TabPFN emerges as a strong classifier with performance better (2x) or on par (3x) compared to AutoGluon$^+$. This confirms the results seen on ML benchmark~\cite{hollmann_tabpfn_2023} and is in itself an impressive result. In particular, TabPFN can leverage its diverse prior to have leading performance ($F_1=0.68$) in the strongly imbalanced Welded Beam dataset. AutoGluon$^+$ shows, as expected, very high performance across the board and is the leading method on the remaining three datasets. Conversely, older machine learning methods (SVM, KNN and DT) are overall clearly behind, except on the Welded Beam (+) dataset where samples are intelligently located.

Considering the Solar HEX, Welded Beam, and Welded Beam (+) datasets, Fig.~\ref{fig:f1_boxplot_full} also demonstrates that TabPFN and XGBoost are sample-order invariant methods. Methods that rely upon hyperparameter tuning naturally introduce additional stochasticity that can lead to different models even if the same dataset is provided. It is also noteworthy that hyperparameter tuning alone---i.e., without ensembling and bagging, can lead to overfitting despite the use of cross-validation in these small datasets. Indeed, XGBoost with the default parameters is better or on par with the tuned variant XGBoost$^\ast$ on seven datasets.

\subsection{Performance with Partial Datasets and Data Efficiency}
To go further, Table~\ref{tab:mean_f1} provides the mean $F_1$ score and standard deviation achieved when considering only 10\% and 50\% of the full dataset. Similarly to the results for the full datasets, TabPFN has leading performance on four and five datasets, respectively. Other methods that get the best mean performance on some datasets include AutoGluon and XGBoost. A more complete picture of the change in performance as dataset size increases is provided in Fig.~\ref{fig:f1-learning-curve} in the Appendix, and confirms the trend described so far. It is also worth mentioning that the observed trends hold when considering the accuracy metric.\footnote{All the results can be obtained and analyzed further: \url{https://decode.mit.edu/projects/pfns4ed/}}

\begin{table*}
    \small 
    \centering
    \caption{Mean and standard deviation of the $F_1$ scores of each method for each trainset when considering 10\% and 50\% of the available dataset. The highest mean value for each dataset is highlighted in blue. We observe that TabPFN has the highest mean performance in 4 out of 8 and 5 out of 8 problems for these trainset sizes.}
    \label{tab:mean_f1}
 \textbf{10\% of the trainsets}\vspace{1mm}
    \resizebox{\textwidth}{!}{ %
\begin{tabular}{@{}l*{8}{c}@{}} %
    & \makecell{Airfoil\\binary} & \makecell{Airfoil\\multi-class} & \makecell{FRAMED\\validity} & \makecell{FRAMED\\safety} & Solar HEX & \makecell{3-bar\\truss} & Welded beam & \makecell{Welded\\beam (+)} \\
    \midrule
    TabPFN & \textcolor{blue}{$0.685 \pm 0.073$} & \textcolor{blue}{$0.468 \pm 0.039$} & $0.514 \pm 0.076$ & $0.510 \pm 0.017$ & \textcolor{blue}{$0.930 \pm 0.028$} & $0.529 \pm 0.184$ & $0.094 \pm 0.122$ & \textcolor{blue}{$0.840 \pm 0.045$} \\
    XGBoost & $0.606 \pm 0.068$ & $0.461 \pm 0.043$ & $0.418 \pm 0.088$ & \textcolor{blue}{$0.573 \pm 0.028$} & $0.894 \pm 0.034$ & \textcolor{blue}{$0.638 \pm 0.036$} & $0.049 \pm 0.066$ & $0.696 \pm 0.064$ \\
    AutoGluon$^+$ & $0.642 \pm 0.075$ & $0.445 \pm 0.071$ & \textcolor{blue}{$0.535 \pm 0.069$} & $0.555 \pm 0.032$ & $0.901 \pm 0.041$ & $0.595 \pm 0.084$ & $0.093 \pm 0.088$ & $0.804 \pm 0.044$ \\
    AutoGluon$^\sim$ & $0.576 \pm 0.095$ & $0.448 \pm 0.056$ & $0.493 \pm 0.075$ & $0.525 \pm 0.030$ & $0.877 \pm 0.062$ & $0.535 \pm 0.114$ & \textcolor{blue}{$0.108 \pm 0.126$} & $0.726 \pm 0.114$ \\
    KNN$^\ast$ & $0.524 \pm 0.118$ & $0.373 \pm 0.078$ & $0.459 \pm 0.123$ & $0.379 \pm 0.029$ & $0.865 \pm 0.042$ & $0.542 \pm 0.101$ & $0.104 \pm 0.139$ & $0.752 \pm 0.060$ \\
    SVM$^\ast$ & $0.661 \pm 0.063$ & $0.462 \pm 0.063$ & $0.447 \pm 0.092$ & $0.472 \pm 0.036$ & $0.913 \pm 0.027$ & $0.562 \pm 0.156$ & $0.057 \pm 0.105$ & $0.799 \pm 0.052$ \\
    XGBoost$^\ast$ & $0.593 \pm 0.074$ & $0.434 \pm 0.044$ & $0.480 \pm 0.108$ & $0.551 \pm 0.029$ & $0.903 \pm 0.036$ & $0.606 \pm 0.056$ & $0.050 \pm 0.073$ & $0.679 \pm 0.067$ \\
    DT$^\ast$ & $0.433 \pm 0.151$ & $0.344 \pm 0.074$ & $0.281 \pm 0.156$ & $0.304 \pm 0.073$ & $0.853 \pm 0.063$ & $0.472 \pm 0.173$ & $0.027 \pm 0.070$ & $0.653 \pm 0.081$ \\
\end{tabular}

    }\vspace{3mm}

 \textbf{50\% of the trainsets}\vspace{1mm}
    \resizebox{\textwidth}{!}{ %
\begin{tabular}{@{}lllllllll@{}}
    & \makecell{Airfoil\\binary} & \makecell{Airfoil\\multi-class} & \makecell{FRAMED\\validity} & \makecell{FRAMED\\safety} & Solar HEX & \makecell{3-bar\\truss} & Welded beam & \makecell{Welded\\beam (+)} \\
    \midrule
TabPFN & \textcolor{blue}{$0.777 \pm 0.027$} & \textcolor{blue}{$0.604 \pm 0.033$} & $0.695 \pm 0.038$ & $0.582 \pm 0.027$ & \textcolor{blue}{$0.980 \pm 0.004$} & \textcolor{blue}{$0.834 \pm 0.015$} & \textcolor{blue}{$0.582 \pm 0.067$} & $0.904 \pm 0.011$ \\
XGBoost & $0.727 \pm 0.034$ & $0.589 \pm 0.038$ & $0.697 \pm 0.044$ & $0.644 \pm 0.020$ & $0.968 \pm 0.007$ & $0.811 \pm 0.014$ & $0.275 \pm 0.088$ & $0.824 \pm 0.013$ \\
AutoGluon$^+$ & $0.756 \pm 0.041$ & $0.584 \pm 0.039$ & \textcolor{blue}{$0.758 \pm 0.045$} & \textcolor{blue}{$0.653 \pm 0.026$} & $0.975 \pm 0.007$ & $0.798 \pm 0.020$ & $0.378 \pm 0.097$ & \textcolor{blue}{$0.905 \pm 0.013$} \\
AutoGluon$^\sim$ & $0.727 \pm 0.042$ & $0.567 \pm 0.043$ & $0.714 \pm 0.070$ & $0.624 \pm 0.023$ & $0.962 \pm 0.016$ & $0.775 \pm 0.030$ & $0.286 \pm 0.160$ & $0.871 \pm 0.041$ \\
KNN$^\ast$ & $0.657 \pm 0.062$ & $0.494 \pm 0.043$ & $0.594 \pm 0.066$ & $0.437 \pm 0.024$ & $0.963 \pm 0.007$ & $0.729 \pm 0.021$ & $0.254 \pm 0.150$ & $0.860 \pm 0.015$ \\
SVM$^\ast$ & $0.752 \pm 0.036$ & $0.540 \pm 0.038$ & $0.649 \pm 0.041$ & $0.548 \pm 0.035$ & $0.967 \pm 0.005$ & $0.794 \pm 0.019$ & $0.023 \pm 0.071$ & $0.896 \pm 0.008$ \\
XGBoost$^\ast$ & $0.718 \pm 0.043$ & $0.556 \pm 0.044$ & $0.752 \pm 0.037$ & $0.637 \pm 0.023$ & $0.967 \pm 0.007$ & $0.802 \pm 0.017$ & $0.187 \pm 0.117$ & $0.820 \pm 0.012$ \\
DT$^\ast$ & $0.590 \pm 0.050$ & $0.435 \pm 0.050$ & $0.632 \pm 0.117$ & $0.470 \pm 0.069$ & $0.955 \pm 0.012$ & $0.633 \pm 0.062$ & $0.021 \pm 0.065$ & $0.783 \pm 0.027$ \\
\end{tabular}
}

\end{table*}

\begin{table*}[ht]
    \centering
    \caption{Relative average data efficiency to attain 90\% of the best performance to the first method reaching the threshold. The method with the highest efficiency is highlighted in blue. We observe that TabPFN achieves the highest data efficiency in 6 out of 8 datasets, also achieving the highest average data efficiency, closely followed by the Autogluon models.}
    \label{tab:data-efficiency}
\begin{tabular}{@{}l*{9}{r}@{}}
    & \makecell{Airfoil\\binary} & \makecell{Airfoil\\multi-class} & \makecell{FRAMED\\validity} & \makecell{FRAMED\\safety} & Solar HEX & \makecell{3-bar\\truss} & Welded beam & \makecell{Welded\\beam (+)} & \makecell{Average}\\ \midrule
TabPFN & \textcolor{blue}{95.6\%} & \textcolor{blue}{85.5\%} & 7.0\% & 52.4\% & \textcolor{blue}{99.0\%} & \textcolor{blue}{96.5\%} & \textcolor{blue}{100.0\%} & \textcolor{blue}{99.6\%} & \textcolor{blue}{79.4\%} \\
XGBoost & 45.7\% & 71.5\% & 84.8\% & 48.8\% & 95.4\% & 89.1\% & 0.0\% & 44.1\% & 59.9\% \\
AutoGluon$^+$ & 78.3\% & 53.7\% & \textcolor{blue}{91.8\%} & \textcolor{blue}{96.0\%} & 97.0\% & 82.2\% & 1.0\% & 91.1\% & 73.9\% \\
AutoGluon$^\sim$ & 66.8\% & 60.1\% & 50.2\% & 77.3\% & 95.3\% & 70.9\% & 12.3\% & 80.1\% & 64.1\% \\
KNN$^\ast$ & 13.9\% & 3.8\% & 0.0\% & 0.0\% & 92.0\% & 0.1\% & 0.0\% & 76.5\% & 23.3\% \\
SVM$^\ast$ & 81.3\% & 27.0\% & 0.0\% & 15.5\% & 98.3\% & 72.9\% & 0.0\% & 92.0\% & 48.4\% \\
XGBoost$^\ast$ & 49.2\% & 47.8\% & 70.6\% & 90.8\% & 96.3\% & 83.3\% & 0.0\% & 36.0\% & 59.3\% \\
DT$^\ast$ & 0.0\% & 0.0\% & 0.0\% & 42.7\% & 92.8\% & 0.0\% & 0.0\% & 3.4\% & 17.4\% \\
\end{tabular}
\end{table*}

Given the good performance with only fractions of the datasets, a natural extension is to consider data efficiency. Table \ref{tab:data-efficiency} offers a detailed assessment of the relative data efficiency of various methods to achieve 90\% of their best performance. The data efficiency is computed relative to the first method that reaches this performance threshold in each dataset and for each run.

TabPFN emerges as the most data-efficient method across the board. Out of the eight datasets presented, TabPFN achieves the highest data efficiency in six. It showcases its superiority, especially in the binary classification of Airfoil (95.6\%), multi-class classification of Airfoil (85.5\%), and in the datasets of Solar HEX (99.0\%), Three-bar truss (96.5\%), and both variations of the Welded beam (100.0\% and 99.6\% respectively). Overall, it holds the highest average data efficiency of 79.4\%.
While TabPFN is the most data-efficient on average, the AutoGluon models give stiff competition in certain datasets. Specifically, AutoGluon$^+$ excels in the FRAMED safety (91.8\%) and FRAMED validity (96.0\%) datasets, outperforming TabPFN. Moreover, while TabPFN leads in average efficiency, the difference between TabPFN and AutoGluon$^+$ is relatively narrow, with the latter having an average efficiency of 73.9\%.

In terms of performance variability, the table highlights the variable efficiency of some methods across different datasets. For instance, KNN$^\ast$ demonstrates a striking efficiency of 92.0\% in the Solar HEX dataset but drops to a meager 0.1\% in the Three-bar truss dataset. Similar trends can be observed for models like SVM$^\ast$ and XGBoost$^\ast$. It's notable that certain methods, specifically DT$^\ast$ and KNN$^\ast$, generally struggle in terms of data efficiency across the majority of datasets, with averages of 17.4\% and 23.3\% respectively.

In terms of problems, the Welded beam dataset proves to be especially polarizing. On one hand, TabPFN achieves a perfect score of 100.0\% efficiency, meaning that regardless of data split, it is the first method to reach the performance threshold.  On the other hand, several models, including XGBoost, KNN$^\ast$, and DT$^\ast$, never reach the performance threshold, resulting in a 0.0\% efficiency.

Overall, while TabPFN asserts its dominance in data efficiency in the majority of datasets, it's crucial to consider the specificities of each dataset when choosing a model, as some might offer competitive or even superior performance in specific contexts.

\subsection{Overall Method Ranking}
\begin{figure}[ht]
    \centering
    \includegraphics[width=\linewidth]{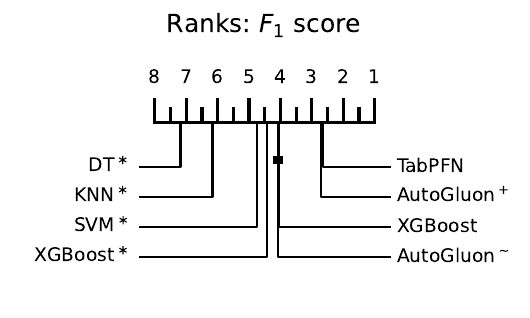}
    \includegraphics[width=\linewidth]{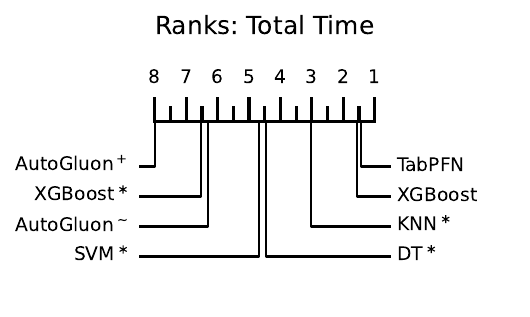}
    \caption{Overall results: Critical difference plot on average ranks in terms of $F_1$ score (top) and total time (bottom) across datasets and splits with a Wilcoxon significance analysis (Holm's adjustment for multiple comparisons). Smaller ranks are better and statistically indistinguishable methods are connected with a black bar. We observe that TabPFN has the lowest rank in both precision and time.
    }
    \label{fig:critical-plots}
\end{figure}

Combining all the data splits and dataset fractions, we perform a statistical analysis to compare the methods based on $F_1$ score and total time. The results from our statistical analysis, shown as critical difference plot in Fig.~\ref{fig:critical-plots},  confirm that TabPFN achieves the best $F_1$ score across the board (rank of 2.65), closely followed by AutoGluon$^+$ (rank of 2.70). While close, the large number of conducted experiments allows for this difference to be statistically significant given the selected 0.05 threshold. Further, TabPFN is also the fastest method (rank of 1.44), followed by XGBoost (rank of 1.56).  As expected, AutoGluon$^+$ is the slowest method (rank 8.00) due to training more models and exploring more hyperparameters.

Apart from  TabPFN and XGBoost which have low scores on both metrics, the other methods follow the more expected performance-speed trade-off.

\subsection{Per Dataset Performance-Speed Trade-Off}
To investigate the performance-speed trade-off, we consider both metrics in a multi-objective fashion. First, we aggregate the performance of each method across a dataset by considering the area under the curve (AUC) formed by the $F_1$ score or total time as a function of the trainset fraction for each data split. Then, considering that $F_1$ should be maximized while the total time should be minimized, we compute for each data split the Pareto rank of each method.

\begin{figure*}[ht]
    \centering
    \includegraphics[width=\linewidth]{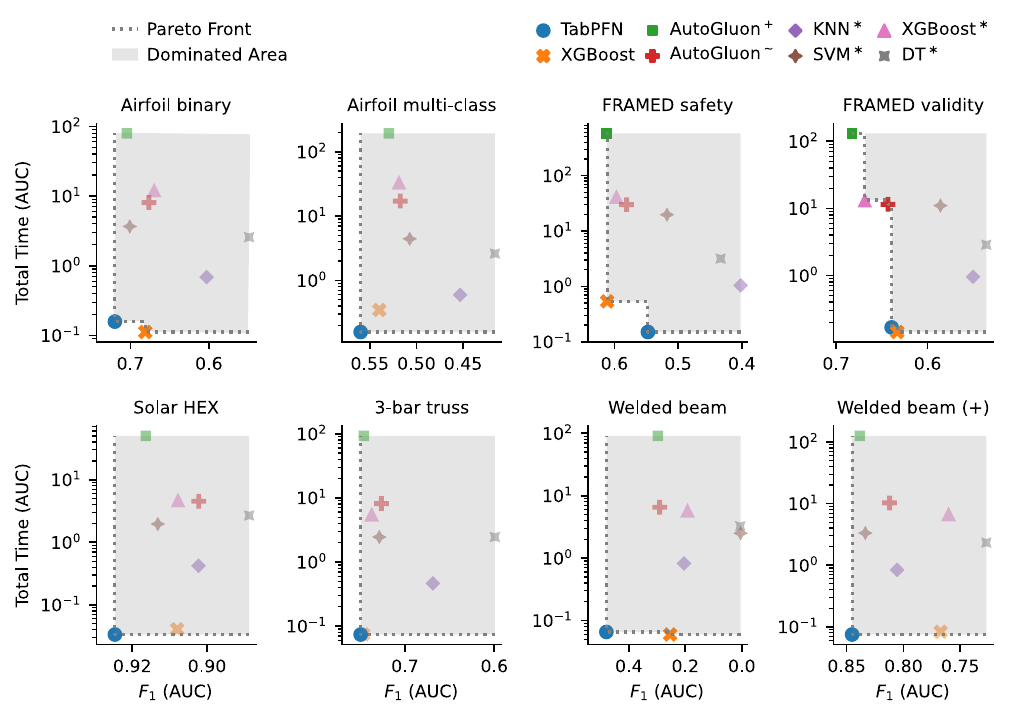}
    \caption{Trade-off between performance and total execution time (log-scale) for each method and dataset. The performance axis is reverted so that the ideal method is in the lower left corner. The resulting Pareto front is shown as a gray dotted line and dominated points are in pastel. We observe that TabPFN is always on the Pareto front.}
    \label{fig:perf_speed_auc}
\end{figure*}

Using this approach, we find that TabPFN has an average Pareto rank of 1.0 across the 20 data splits for all eight datasets, showing that it presents as a recommended choice for both speed and performance. It is closely followed by XGBoost, with an average rank of 1.375. Dataset-specific details are shown in Fig.~\ref{fig:perf_speed_auc}. With this detailed view it becomes clear, that TabPFN is either leading in both $F_1$ score and total time or, when TabPFN is less accurate, e.g., for the FRAMED datasets, it is part of the performance-speed non-dominated front.

\section{Discussion}

\paragraph{Understanding the Complexity of Design Classification Problems}
Analyzing the eight selected design problems provides a clearer understanding of their inherent difficulty. We utilize two specific metrics to gauge this. Firstly, by ranking the problems based on the collective F1 scores achieved by all eight ML methods, we discern that the Welded beam problem is the most challenging, given its lowest average performance. In contrast, the Solar HEX problem seems to be the easiest, as it commands the highest average performance. The second metric, the variance in performance across methods, reveals the criticality of method selection for specific domains. Solar HEX showcases minimal variance, suggesting that the choice of method doesn't significantly affect its performance outcome. However, for problems like Welded beam and FRAMED safety, the considerable variance underscores the crucial role of method selection in determining the results.

\paragraph{Recommendations for Engineering Design Classification Methods}
While previous work on classification for design~\cite{sharpe_comparative_2019} found that there are no clear recommendations for which classical classification model to choose for engineering design, our results investigating the latest ML methods led us to make the following recommendations. TabPFN is, overall, both the most accurate and the fastest method currently available. It also has the best average data-efficiency and should as such be the method of choice for datasets that fit its limitations (less than 100 features and less than 5000 data points). The second method of choice, if training time is not an issue, is AutoGluon with \textit{best} or \textit{medium} quality presets. Finally, XGBoost with default settings, while being less accurate in general, still offers a good performance-time ratio. We hope that the clearer guidelines on method selection support engineers who wish to use data-driven methods for their applications.

\begin{figure*}
    \centering
    \includegraphics[width=\linewidth]{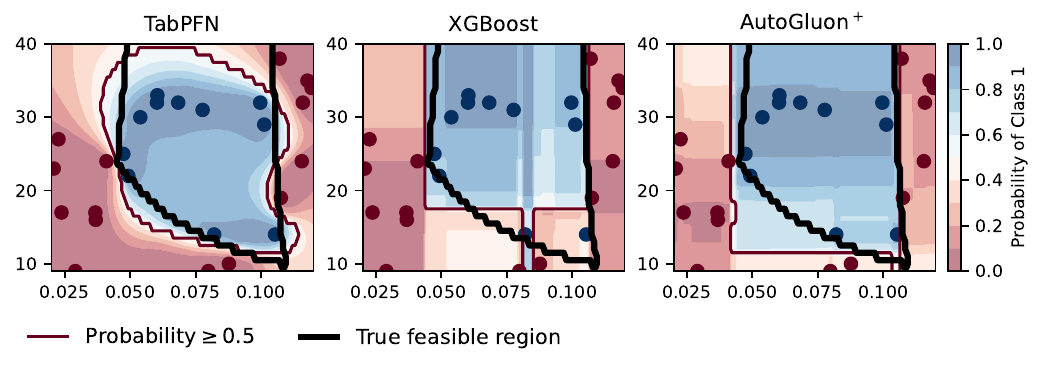}
    \caption{Probability predictions for class 1 (feasible designs) for the three overall best-performing models for the two-dimensional Solar HEX problem. The 5\% train set is used (25 samples) and a grid of 100x32 is used to create the heatmap.}
    \label{fig:uncertainty}
\end{figure*}

\paragraph{Uncertainty Estimates and Differentiability}
We also want to highlight additional properties of TabPFN in particular and prior-data fitted networks in general. TabPFN supports classification problems with up to ten classes and can provide well-calibrated probabilities for each of them~{\cite{muller_transformers_2022}}. To illustrate what it means, Fig.~{\ref{fig:uncertainty}} qualitatively compared the probability of feasibility for the two-dimensional Solar HEX problem obtained from the overall three best-performing models: TabPFN, XGBoost, and AutoGluon$^+$. Contrary to the latter two which are based on decision trees, TabPFN provides smoother transitions in line with typical Bayesian models. Well-calibrated probability predictions could be especially beneficial for the engineering design community: when the uncertainty is high, additional simulations or different models could be used. Further, since PFNs are deep neural networks, they are completely differentiable and their gradient can be used to perform gradient-based optimization to solve the inverse design problem: finding a design that achieves a certain class.

\subsection{Limitations}
This study, while offering significant insights into the use of PFN models for engineering design challenges, carries some inherent limitations. Firstly, TabPFN's performance was gauged predominantly on its speed and accuracy; however, in the complex realm of engineering design, other metrics like interpretability might be equally crucial. The eight datasets employed, although representative, might not encompass the full spectrum of variability and intricacy present in real-world engineering scenarios. We hope that our work motivates other researchers to use these methods for their engineering design problems, while also benchmarking new algorithm developments on the set of problems we used.
We also haven't exhaustively evaluated the influence of specific preprocessing techniques on TabPFN's efficacy.
Lastly, while TabPFN exhibited commendable performance, it might need to be complemented with domain-specific knowledge in many intricate engineering design challenges to realize its full potential. 

\subsection{Future Work}

Our findings in this paper pave the way for numerous future research directions. Key areas of exploration include leveraging larger test datasets and a deeper dive into feature engineering to uncover more intricate relationships within the design data. Given the speed of PFN models, the potential of integrating these classifiers within an active learning framework, which focuses on selecting the most informative samples, is an exciting prospect. Similarly, their speed lends itself well to their use within ensembles. Additionally, adapting the approach to regression tasks can broaden its utility. 
Considering the \textit{emergent abilities} described for other transformer models, studying the effect of model size on performance seems a natural extension. Lastly, the incorporation of uncertainty quantification methods will enhance the reliability and robustness of our models, ensuring they not only predict but also gauge the confidence of their predictions.

\section{Conclusion}
In this work, we have evaluated a game-changing approach to machine learning that leverages prior-data fitted networks to remove the need for domain-specific training for classification tasks. To support the rigorous evaluation, we have defined eight diverse problems and generated corresponding datasets. Based on those, we evaluated eight data-driven classification methods: classical approaches (KNN, SVM, and DT), an ensembling method (XGBoost), an AutoML algorithm (AutoGluon), and a prior-data fitted network (TabPFN). Across datasets, TabPFN was overall the most accurate, data-efficient, and fastest approach, followed by AutoGluon for accuracy and XGBoost for speed.

As such TabPFN stands out as a broadly applicable classifier that does not require machine learning expertise to tune hyperparameters, provides well-calibrated uncertainty estimates, and is differentiable. All of which make it an excellent candidate to become a standard for data-driven tasks in engineering design. This is especially striking since it has been trained exclusively on synthetic data, yet seemingly encompassing engineering data distributions. In contrast, other models that require training from scratch can offer better customization and domain-specific performance when the necessary time, resources, and expertise are available.

Pre-trained transformer models that use in-context learning for tabular data open up a new field of possibilities within engineering to promote fast, accurate, and easy-to-use data-driven methods to support the design process in general, and in industry in particular. 
Future work directions include extending this work by looking into their use for Bayesian optimization, as well as, the development of subdomain-specific pre-trained networks for an even higher data efficiency.

\section*{Acknowledgment} %
We would like to thank Dr. Carolyn Seepersad, Dr. Tyler Wiest, and their co-authors for sharing their data and code for the Solar HEX, 3-bar truss, and Welded Beam benchmark problems.

\section*{Funding Data}

\begin{itemize}
\item The Swiss National Science Foundation (Postdoc.mobility No.\ P500PT\_206937).
\end{itemize}

\appendix   %
\section{Additional Results}

\begin{figure*}
    \centering
    \includegraphics[width=\linewidth]{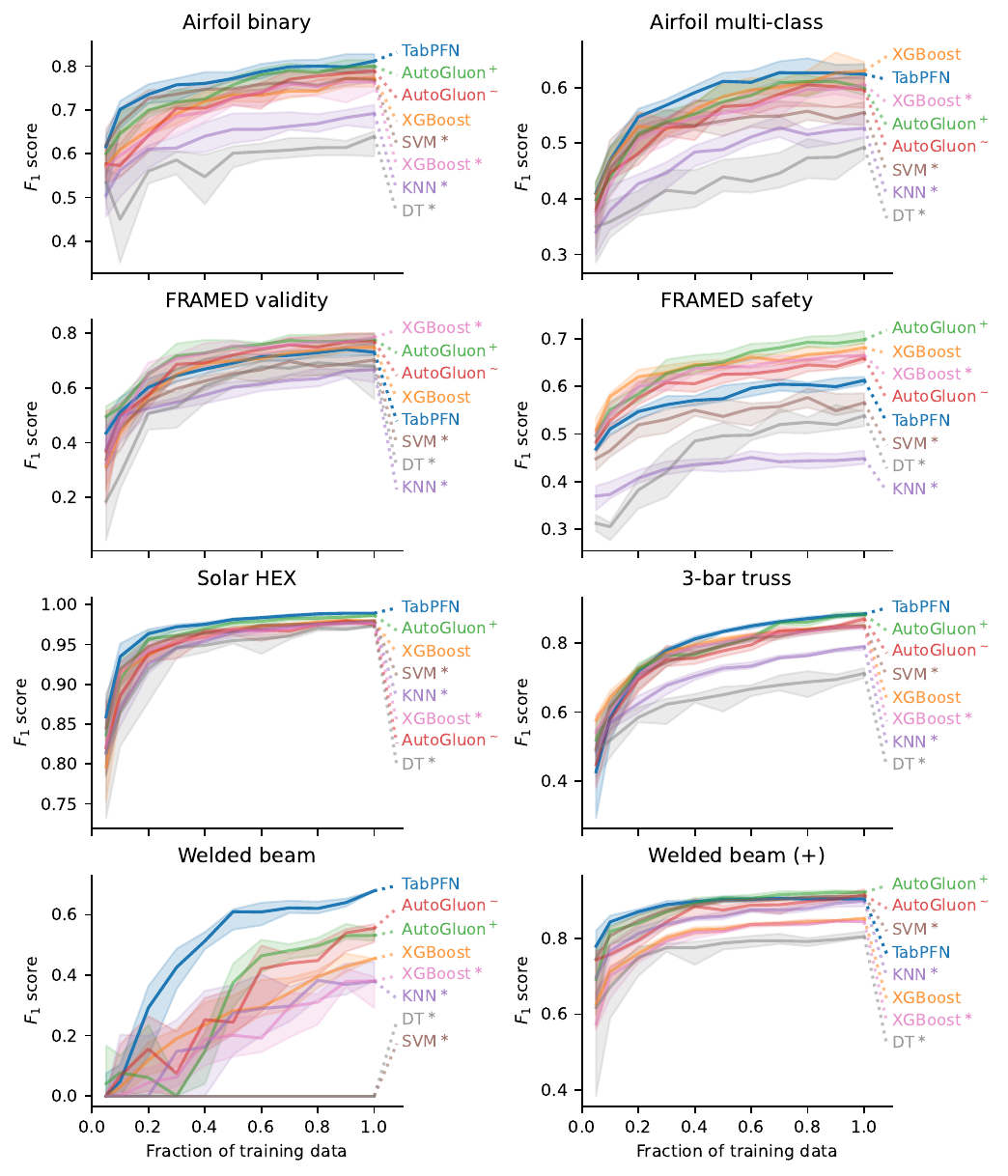}
    \caption{Median and 95\% confidence interval of the $F_1$ score as a function of the fraction of training data supplied.}
    \label{fig:f1-learning-curve}
\end{figure*}

All the performance results can be summarized in a set of \textit{learning curves} showing how the performance changes with increased data, see Fig.~\ref{fig:f1-learning-curve}.

\bibliographystyle{asmejour}   %

\bibliography{src/bibliography} %

\end{document}